\documentclass[prb,aps]{revtex4}

\usepackage{graphicx}
\usepackage{dcolumn}
\usepackage{amsmath}

\newcommand{\beq}{\begin{equation}}
\newcommand{\eeq}{\end{equation}}

\begin{document}

\title[Short Title]{Equilibrium Contact Probabilities in Dense Plasmas}

\author{B. Militzer}
\affiliation{Geophysical Laboratory, Carnegie Institution of Washington,
             5251 Broad Branch Road, NW,
             Washington, DC 20015} 
\author{E.L. Pollock}
\affiliation{Lawrence Livermore National Laboratory,
         University of California, Livermore, CA 94550}

\begin{abstract}
Nuclear reaction rates in plasmas depend on the overlap (contact)
probability of the reacting ions.  Path integral Monte Carlo
(PIMC) calculations are used here to determine these contact
probabilities, $g(0)$, for the one component plasma (OCP) with
emphasis on many-body quantum effects which can lead to order of
magnitude changes. An intuitive explanation for these effects is
presented.  The small $r$ behavior of $g(r)$ for quantum systems
and the relation to free energies is then derived and compared to
the path integral results.  Going beyond the uniform background
approximation, electron screening effects and the limits of the
``constant energy shift'' approximation are discussed.
Thermodynamic properties for the quantum OCP are analyzed in a
final section.

\end{abstract}

\date{\today }

\pacs{??.?? ??.?? ??.?? }

\maketitle

\section{Introduction}

      Due to its importance in calculating nuclear reaction rates, the
small $r$ behavior of the radial distribution function, $g(r)$, in
ionized systems has been a subject of major interest since the original 
work of Salpeter~\cite{salpeter}. For the uniform background model of a
classical plasma, simulation studies have produced a quantitative picture
of the role of many-body effects~\cite{graboske, slattery}. Changes
due to quantum effects are less thoroughly 
understood~\cite{ogata1, ogata2, pollock}.

      This article uses path integral Monte Carlo to compute $g(r)$ and in
particular $g(0)$ for dense plasma models.  Section II presents extensive
results for the quantum one component plasma model of ions in a charge neutralizing
background and gives an intuitive explanation for the trends observed.
For classical systems $g(0)$ has a simple connection to a free energy
difference. The relation is more complicated in the quantum case and section
III along with the appendix discusses the small $r$ behavior of $g(r)$ in
general. Section IV goes beyond the uniform background approximation to
consider the effects of electron screening~\cite{itoh}. The thermodynamic functions from the
simulations are given in section V.

      The ensuing question of how this information on contact probabilities 
is ultimately used in a reaction rate calculation~\cite{clayton} has not been 
rigorously answered in the literature. 
      The common expression for the equilibrium nuclear reaction rate is obtained
by multiplying the experimentally determined cross section for a particular
reaction, usually as a function of incident momentum, by a particle flux
and averaging over the Maxwellian distribution for the
relative incident momentum. Multiplying by the target density gives 
the reaction rate per volume between species 1 and 2 at temperature $k_{B}T=1/\beta$
\beq
{\cal R}_{12} =n_{1}n_{2}\int \sigma(p)  
     \left( \begin{array}{c} \underline{ p }\\ \mu \end{array} \right)
      \left(\frac{\beta}{2\pi\mu}\right)^{3/2}
 e^{-\beta p^{2}/2\mu} d^3p\;.
                                          \label{eq1.1}
\eeq
For charged particle reactions, the cross section $\sigma(p)$ is usually written as
a ``nuclear'' cross section or astrophysical factor times a term representing the
Coulomb barrier penetration. This last step has the advantage of permitting
cross sections measured at accessible experimental energies to be more reliably
extrapolated down
to the lower energies relevant to most astrophysical applications. 

Different prescriptions have been advanced for modifying this reaction rate 
to take into account  many-body ion and electron screening effects. 
The most common prescription is to simply multiply the above reaction rate by
the relative change in the contact probability, $e^{H(0)}$ defined below.
For electron screening effects a constant energy shift (see section IV) in the 
Coulomb barrier penetration term is, however, frequently used.
The work of Brown and Sawyer~\cite{brown-sawyer} perhaps provides a starting 
point for a fuller discussion of this key question.

\section{OCP Contact Probabilities}

      Dense plasma effects on nuclear reaction rates are usually described
in terms of the enhancement in the
contact probability $g(0)$. $g(0)$ is commonly factored into the  two-body 
term, $g_{bin}(0)$, and a term, $H(0)$, representing many-body effects,

\beq
g(r) = g_{bin}(r)e^{H(r)}\;.
           \label{eq2.1}
\eeq
For a classical system $g_{bin}(r)$ is just $e^{-\beta Z_{1}Z_{2}e^{2}/r}$. 
For a quantum system $g_{bin}(r)$ is obtained from the
solution of the Bloch equation for the density matrix.

     For the repulsive Coulomb potential $g_{bin}(0)$ has an analytic expression~\cite{pollock1}
\beq
g_{bin}(0)=(4\pi\beta)^{3/2}\frac{Z^{3}}{2\pi}\int_{0}^{\infty}
   \begin{array}{c}\underline{ k e^{-\beta Z^{2}k^{2}}dk}\\ e^{\pi/k}-1
   \end{array}\;.
           \label{eq2.2}
\eeq
A plot of $g_{bin}(0)$ is shown in the lower panel of Fig. 1. Unlike the classical
$g_{bin}(r)$ which is zero at the origin, the quantum $g_{bin}(0)$ is finite at the
origin and increases with temperature. As expected $g_{bin}(r)$ converges to the
classical result for $r$ larger than the de Broglie thermal wavelength, 
$\lambda_{d}^{2}=\hbar^{2}/2\pi mk_{B}T$.
This is demonstrated in the upper panel of Fig. 1.

\begin{figure}[!]
\includegraphics[angle=-90,width=0.55\textwidth]{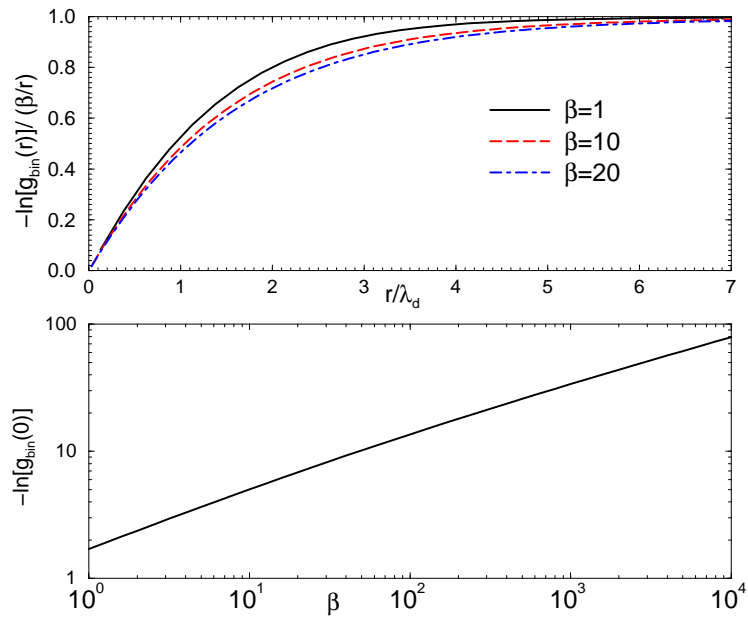}
\caption{
         $g_{bin}(r)$ for the Coulomb potential. The upper panel shows the 
         convergence to the classical limit for $r$ several times the de Broglie
         thermal wavelength, $\lambda_{d}$. The lower panel shows the 
         dependence of $g(0)$ on the temperature. The quantum $g(0)$ 
         is finite, unlike the classical limit.
         }
\label{fig1}
\end{figure}

      In this section, contact probabilities are discussed for the OCP
model consisting of a single ion species in a uniform, 
charge neutralizing background. 
In the classical limit, the nonideal properties of this model depend
only on the dimensionless combination of temperature and density given
by the coupling constant
$\Gamma=Z^{2}e^{2}/k_{B}T a$, where $a$ is the ion sphere radius
defined by $4\pi a^{3} N/V=1$.
$\Gamma$ is a measure of the relative importance of potential to kinetic energy.
Contact probabilities for this classical model have been studied
starting from the Monte Carlo simulations  of Brush, Sahlin 
and Teller~\cite{graboske, brush}.
      Empirically $H(0)$ is dominated by a linear dependence on $\Gamma$ so
it is convenient to define an enhancement factor, $h(0)\equiv H(0)/\Gamma$.

            When quantum effects are included
both density and temperature must be specified and it is
necessary to introduce a quantum parameter $\eta\equiv\Gamma/r_{s}$, 
with $r_{s}=a/a_{0}$, where $a_{0}$ is the Bohr radius for the ions. 
$\eta$ rewritten as $\eta=2\pi\lambda_{d}^{2}/a^{2}$, is seen to be proportional 
to the squared ratio of the de Broglie thermal wavelength
to the ion sphere radius and thus provides an appropriate gauge for
quantum effects.

The many-body $g(r)$ is computed here by averaging over the density matrix 
$e^{-\beta H}$ using Path integral Monte Carlo based on the identity 
\beq
    e^{-\beta H}=\left[e^{-\beta H/M}\right]^{M}
         \label{eq2.4}
\eeq
where $M$ is an arbitrary integer.
Insertion of complete sets of states between the $M$ factors on the right 
hand side of this equation leads to the usual path integral formulation
of the density matrix, written here in real space, 
\beq
\left<{\bf R}|e^{-\beta H}|{\bf R}'\right>\equiv \rho({\bf R},{\bf R}';\beta)=
\int\ldots\int \rho({\bf R},{\bf R}_{1};\tau)\ldots\rho({\bf R}_{M-1},{\bf R}';\tau)
d{\bf R}_{1}\ldots d{\bf R}_{M-1}
     \label{eq2.5}
\eeq
with $\tau=\beta/M$. Each of the $M$ steps in the path now has a high temperature
density matrix $\rho({\bf R}_{k},{\bf R}_{k+1};\tau)$ associated with it. The
integrals are evaluated by Monte Carlo methods. First applied to realistic systems
in reference 12, the implementation details may be found in  recent 
reviews~\cite{cep_rmp}. In the results presented here the high temperature
density matrix was taken as a product of exact pair density matrices. Typical
$M$ values of 10 to 400, depending on $\beta$, gave a discretization error well
below the Monte Carlo statistical uncertainties.

\begin{figure}[!]
\includegraphics[angle=0,width=0.75\textwidth]{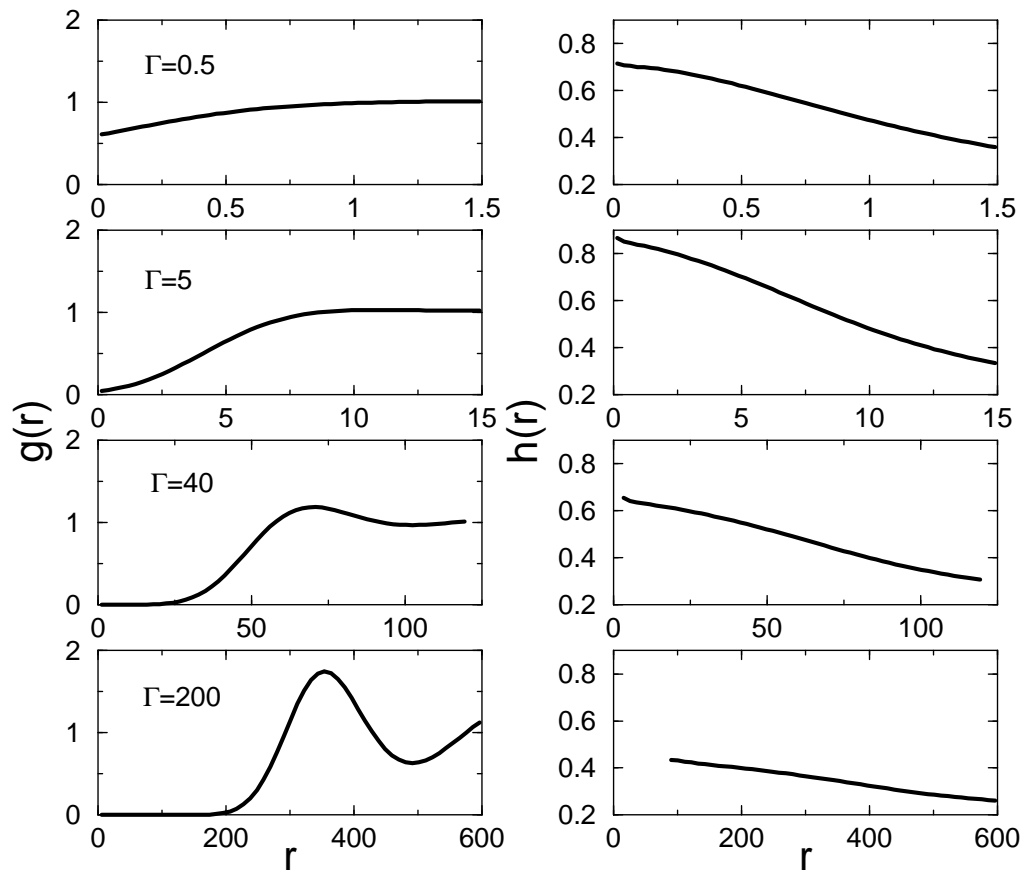}
\caption{
        Pair correlation functions $g(r)$ and the corresponding 
	 many-body enhancement $h(r)$ are shown as a function of the
	 coupling parameter, $\Gamma$, for quantum parameter, $\eta=1$.
        $r$ is in units of the nuclear Bohr radius.
         }
\label{fig2}
\end{figure}

      Typical examples of $g(r)$ and $h(r)=H(r)/\Gamma$
are shown in Fig. 2. Since $g(0)$ is now finite the $h(r)$ curves are considerably
easier to extrapolate to the origin than for the classical OCP. A detailed 
discussion of the formal properties of $g(r)$ at small $r$ is given in section III and
the appendix. $g(0)$, $g_{bin}(0)$ and $h(0)$ are given in table I for a range
of $\Gamma$ and $\eta$ and are displayed in Fig. 3.

\begin{table}[!]
\caption[tab1]{
         Summary of OCP contact probabilities. The coupling parameter $\Gamma = \beta/r_s$
         and quantum parameter $\eta = \Gamma/r_s$, or equivalently $\beta$ and $r_{s}$,
         are listed in columns one through four. The many-body enhancement factor, $h(0)$,
         is defined from the contact probability, $g(0)=g_{bin}(0)e^{H(0)}=
         e^{-P(0)+\Gamma h(0)}$ (columns five through seven).
              }

\vspace*{.1in}
\begin{tabular}{|c|c|r|r|l|l|l|}\hline
$\eta$   & $\Gamma$ & $\beta$ &$r_{s}$&  $-\ln[g(0)]$   & $P(0)$    & $h(0) $      \\\hline
0.1      & 0.5      & 2.5     & 5.0   &  2.205(18)      & 2.638     & 0.87(4)      \\
0.1      & 1.0      & 10      & 10    &  4.06(3)        & 5.014     & 0.95(3)      \\
0.1      & 2.0      & 40      & 20    &  7.22(6)        & 9.243     & 1.01(3)      \\
0.1      & 5.0      & 250     & 50    &  14.61(12)      & 19.77     & 1.03(2)      \\
0.1      & 10       & 1000    & 100   &  23.6(3)        & 33.79     & 1.02(3)      \\
0.1      & 40       & 16000   & 400   &  57.4(14)       & 93.61     & 0.91(4)      \\ \hline
0.25     & 0.5      & 1.0     & 2.0   &  1.292(12)      & 1.707     & 0.82(2)      \\
0.25     & 1.0      & 4.0     & 4.0   &  2.37(3)        & 3.289     & 0.92(3)      \\
0.25     & 2.0      & 16      & 8.0   &  4.25(5)        & 6.192     & 0.97(3)      \\
0.25     & 5.0      & 100     & 20    &  8.59(12)       & 13.50     & 0.98(2)      \\
0.25     & 10       & 400     & 40    &  14.33(3)       & 23.90     & 0.957(3)     \\
0.25     & 40       & 6400    & 160   &  34.3(16)       & 67.32     & 0.83(4)      \\
0.25     & 100      & 40000   & 400   &  59(4)          & 129.5     & 0.71(4)      \\\hline
0.50     & 0.5      & 0.5     & 1.0   &  0.831(7)       & 1.218     & 0.773(14)    \\
0.50     & 1.0      & 2.0     & 2.0   &  1.504(19)      & 2.373     & 0.869(19)    \\
0.50     & 2.0      & 8.0     & 4.0   &  2.68(5)        & 4.530     & 0.92(2)      \\
0.50     & 5.0      & 50      & 10    &  5.53(12)       & 10.17     & 0.93(2)      \\
0.50     & 10       & 200     & 20    &  9.32(14)       & 18.01     & 0.869(14)    \\
0.50     & 40       & 3200    & 80    &  23.1(6)        & 52.45     & 0.734(16)    \\
0.50     & 100      & 20000   & 200   &  40(2)          & 101.3     & 0.61(2)      \\\hline
1.0      & 0.5      & 0.25    & 0.5   &  0.514(4)       & 0.8683    & 0.708(9)     \\
1.0      & 1.0      & 1.0     & 1.0   &  0.917(11)      & 1.704     & 0.787(11)    \\
1.0      & 2.0      & 4.0     & 2.0   &  1.60(3)        & 3.289     & 0.844(14)    \\
1.0      & 5.0      & 25      & 5.0   &  3.29(11)       & 7.540     & 0.85(2)      \\
1.0      & 10       & 100     & 10    &  5.53(19)       & 13.50     & 0.797(19)    \\
1.0      & 40       & 1600    & 40    &  14.8(6)        & 40.31     & 0.637(15)    \\
1.0      & 200      & 40000   & 200   &  41(3)          & 129.5     & 0.442(14)    \\
1.0      & 400      & 160000  & 400   &  63(3)          & 210.0     & 0.369(8)     \\
1.0      & 600      & 360000  & 600   &  83(3)          & 277.7     & 0.324(4)     \\\hline
2.0      & 0.5      & 0.125   & 0.25  &  0.3083(9)      & 0.6175    & 0.6184(18)   \\
2.0      & 1.0      & 0.5     & 0.5   &  0.536(5)       & 1.218     & 0.682(5)     \\
2.0      & 2.0      & 2.0     & 1.0   &  0.923(14)      & 2.373     & 0.725(7)     \\
2.0      & 5.0      & 12.5    & 2.5   &  1.86(6)        & 5.545     & 0.737(13)    \\
2.0      & 10       & 50      & 5.0   &  3.19(16)       & 10.17     & 0.698(16)    \\
2.0      & 40       & 800     & 20    &  9.1(5)         & 30.92     & 0.546(12)    \\
2.0      & 100      & 5000    & 50    &  17.0(8)        & 61.34     & 0.443(8)     \\\hline
\end{tabular}
\end{table}

\begin{figure}[!]
\includegraphics[angle=0,width=0.55\textwidth]{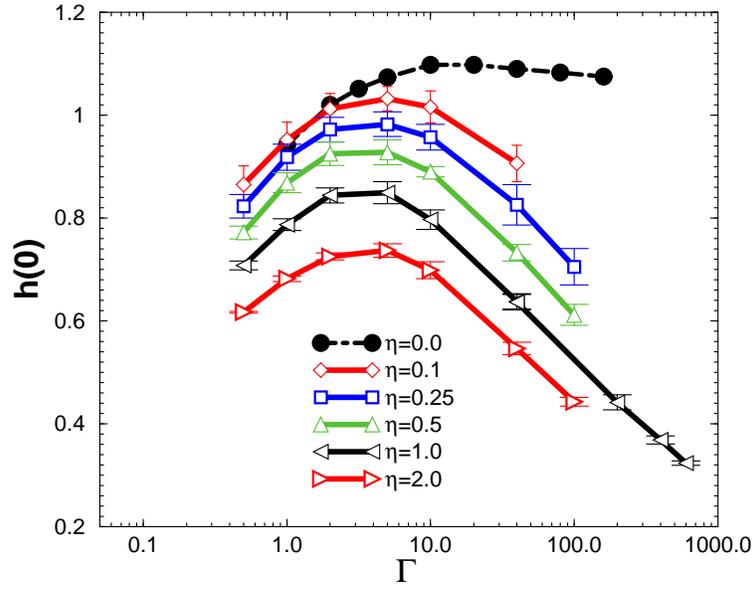}
\caption{
        The many-body enhancement factor $h(0)$ as a function of $\Gamma$
        and quantum parameter $\eta$. Results for the classical OCP, $\eta=0$,
        are from reference~\cite{slattery}.
         }
\label{fig3}
\end{figure}
\newpage

      A clear reduction in $h(0)$ from the classical value (solid circles), 
which becomes more important as $\eta$ increases, is seen in 
Fig.~\ref{fig3}. This yields reaction rates
orders of magnitude smaller than would be predicted by the classical value for $h(0)$
at large $\Gamma$. 

      Fig.~\ref{fig1} provides the basis for  an intuitive 
understanding of this reduction due to quantum effects. $P(r)=-\ln g_{bin}(r)$ can 
be roughly viewed as proportional to an ``effective'' quantum pair potential. 
Increasing $\eta$ means that 
the near neighbors of the reacting pair, located approximately one ion sphere
radius away, are within a de Broglie thermal wavelength. 
Their effective quantum pair potential is then much less than the Coulomb
potential as seen from the top panel of Fig.~\ref{fig1}. 
This reduced repulsion lessens the many-body enhancement for
the quantum system compared to that of a classical system at the same $\Gamma$. 
Reduction from the classical Coulomb value for $h(0)$ is also seen for classical 
screened Coulomb systems~\cite{caillol} and for quantum screened 
Coulomb systems (section IV). The physical explanation is again the reduced
effective repulsion between the reacting pair and its surrounding neighbors
due either to screening or quantum effects or both.
This is a common pattern. A softer effective potential, either from quantum effects
or screening enhances the two body contribution to $g(0)$ but reduces the
many-body contribution.

\begin{figure}[!]
\includegraphics[angle=-90,width=0.55\textwidth]{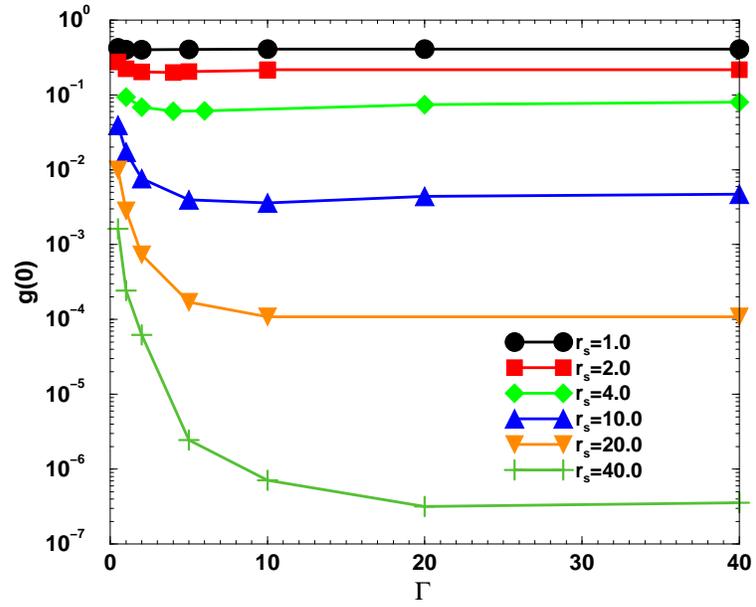}
\caption{
        Contact probabilities, $g(0)$, versus $\Gamma$ for indicated
        $r_{s}$ showing the transition from thermonuclear (strong
        temperature dependence, low $\Gamma$) to pycnonuclear (temperature independent)
        regimes.
        }
\label{fig4}
\end{figure}

      Although $h(0)$ provides a compact way of presenting the data, the full
$g(0)$, shown in Fig.~\ref{fig4} gives a more intuitive, physical picture.
Fig.~\ref{fig4}, shows the transition from
the thermonuclear, temperature dependent regime, at low $\Gamma$ (high temperature)
to the pycnonuclear, density dependent ground state regime at large $\Gamma$.
The low temperature limit seems to be  reached when the de Broglie thermal wavelength
is of the order of $1/3$ the ion sphere radius.
The principle feature of this graph is the smooth behavior of $g(0)$.
$g(0)$ is seen to be a increasing function of the temperature and the density.
For this model there are no peculiar combinations of density and temperature 
where the contact probability, and by implication the reaction rate, has an 
unexpected local maximum.

\newpage

\section{$g(r)$ at small $r$ and relation to free energies}

   The radial distribution function may be expanded as
\beq
    g(r)\equiv {\Omega\over N^{2}}\left<\sum_{i\neq j}\delta ({\bf r}-{\bf r}_{ij})
         \right>= g(0)+Cr^{2}+O(r^{4})\;.
                       \label{eq3.1}
\eeq
In a classical system the coefficient of the $r^{2}$ term is proportional to
mean squared force on the two fused particles. $g(0)$ is related 
to the difference between the free energy of a mixture consisting of the 
fused pair and the N-2 other particles and the free energy of the 
original system\cite{widom}.
For a quantum system the expansion coefficients are more complicated.

     Since the delta function in the definition of $g(0)$ causes the two 
``reacting'' particles to overlap
it might be thought that the relation to the hypothetical mixture free energy which holds
in the classical case would be sufficient in general. For those comfortable with
the path integral ideas used in this paper the cartoon in Fig. 5
may give an intuitive understanding of why this is not true. Others are relegated
to the detailed derivation in the appendix.

\begin{figure}[!]
\includegraphics[angle=-90,width=0.60\textwidth]{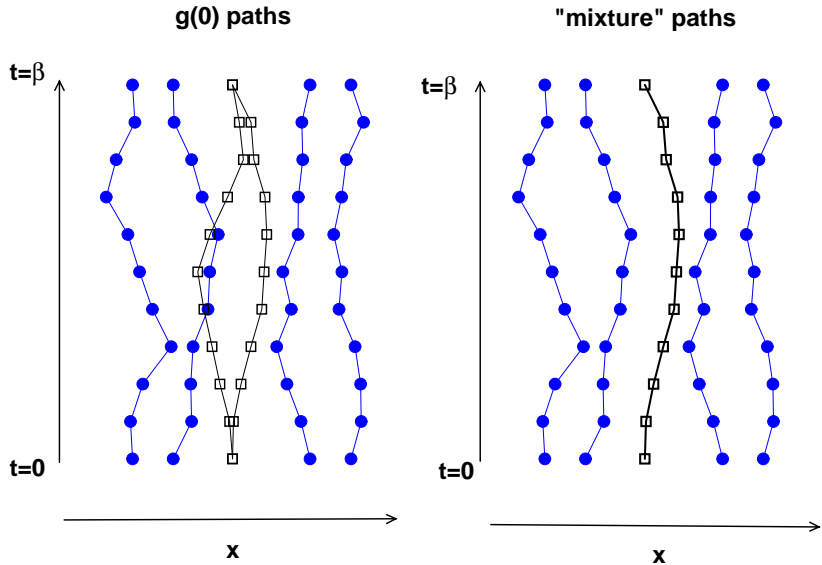}
\caption{
         Intuitive illustration of the relationship between $g(0)$ and the 
         ``mixture'' free energy for a quantum system and its classical limit.
         The left panel shows a typical set of  discretized (10 steps) paths 
         which would enter the calculation of $g(0)$ for a six particle 
         system. The nodes for the paths of the two ``reacting'' particles
         (in center) are shown as open squares. Nodes of the surrounding particle
         paths are shown as filled circles. Because of the delta function in 
         the definition of $g(0)$ the two paths for the reacting pair overlap
         at imaginary times $t=0$ and $\beta$. The right panel shows typical 
         paths for the density matrix,
         integrated to get the free energy, of a five particle mixture where the
         particle in center (nodes shown as open squares) has a mass and charge equal
         to the combined mass and charge of the reacting particles in the left panel. 
         In the classical limit, when the de Broglie thermal wavelength goes to zero, the
         paths for the reacting pair (left panel) or mixture particle (right panel)
         become straight vertical lines. The
         potential energy at the reacting pair or mixture particle 
         due to the four neighbors is then the same for both systems implying
         that $g(0)$ can be derived from the free energy difference.
        }
\label{fig5}
\end{figure}

            The two panels of this cartoon show the 
paths (discretized here into 10 segments) that contribute to $g(0)$ 
for a six particle system (left panel) and the paths contributing to the 
density matrix of a
mixture of five particles (right panel) where the two ``reacting'' particles 
have been fused into one particle with the combined charge 
and mass of the reacting particles. In the figure the nodes in the discretized paths of
the reacting (or fused) pair are shown as open squares. The obvious difference in the paths 
is that in the $g(0)$ case the delta function in the definition causes the 
paths of the reacting particles to overlap at imaginary times $t=0$ and $t=\beta$
but not at other times.
In the mixture, by contrast, there is only one path for the fused particle. 
Since this fused particle has the combined mass of the reacting pair 
it  would be a ``more classical'' particle and typically have a less fluctuating path.

     The scale for the deviation of these paths from straight vertical lines is set by 
the de Broglie thermal
wavelength. The different contribution from the two cases (left and right panel)
comes from the potential energy around the paths of the reacting (or fused) pair due 
its neighbors.
In the classical limit the de Broglie thermal wavelength goes to zero and the paths 
reduce to lines. The potential energy of the other particles at the reacting pair 
(left panel) is then the same as for the fused particle in the mixture case (right panel).
Hence the correspondence in the classical limit.
This also indicates that the classical relation between $g(0)$ and the free energy
difference holds when the reacting particles, but not necessarily the other neighbors,
are treated classically.

      Details for the $g(r)$ expansion are given in the appendix. It follows the
original work of Jancovici\cite{jancovici} and Alastuey and 
Jancovici\cite{alastuey}. We differ in explicitly separating the $g_{bin}(0)$
term, which is given numerically by Eq.~\ref{eq2.2}, and also the free energy 
difference, approximations for which may be available from other theories.
This involves a slight rearrangement of terms.

To lowest order in the quantum parameter $\eta$ and $r$,
\beq
    H(r)=-\beta [F(1,N-2)-F(0,N)]-{\Gamma\over 4a^{2}}\left<r^{2}\right>
                            \label{eq3.2}
\eeq
where the first argument in the interaction free energies, $F$, denotes the number of
combined mass, combined charge ions. $\left<r^{2}\right>$ is calculated from the relative pair
density matrix
\beq
\left<r^{2}\right>(r,\beta)\equiv
    \frac{1}{\beta}\int_{0}^{\beta}ds\int d{\bf r}'
  \begin{array}{c}
   \underline{ \rho({\bf r},{\bf r}';\beta-s) r'^{2}
                              \rho({\bf r}',{\bf r};s)}\\
         \rho({\bf r},{\bf r};\beta)
   \end{array}\;.
                                   \label{eq3.3}
\eeq

In the classical limit $\left<r^{2}\right>=r^{2}$ and the usual result
\beq
 H(r)=-\beta [F(1,N-2)-F(0,N)]-{\Gamma\over 4a^{2}}r^{2}
                            \label{eq3.4}
\eeq
is regained. Using the Debye-H{\"u}ckel free energies~\cite{landau} in Eq.~\ref{eq3.4}
gives the classical weak coupling limit, $h(0)=\sqrt{3\Gamma}$.   
As suggested in Fig. 6 the classical values will approach this for $\Gamma\leq 0.1$.
Although simulation results for the classical $h(0)$ at $\Gamma<1$ have not been
published the approach to the Debye-H{\"u}ckel limit is clearly seen in calculations
based on the HNC approximation~\cite{dewitt3}.

\begin{figure}[!]
 \includegraphics[angle=-90,width=0.55\textwidth]{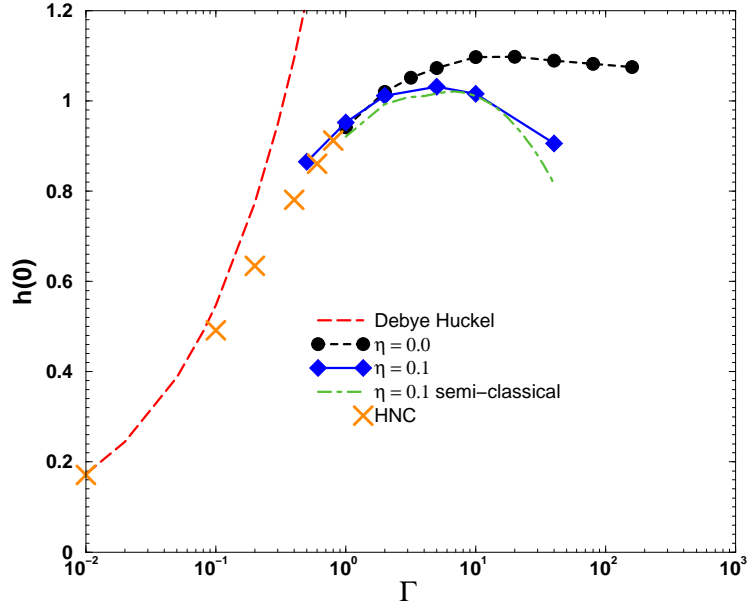}
\caption{ Debye-H{\"u}ckel approximation for $h(0)$ (dashed line) as low $\Gamma$
          limit for the classical, $\eta=0$, system. PIMC (solid diamonds)
          and semiclassical values (dot-dashed line), Eq. 10, for $h(0)$ 
          at $\eta=0.1$ are also shown. Hypernetted chain integral equation
          results (crosses) are from Ref. 18
            }
\label{FIG6}
\end{figure}

      More generally Eq.~\ref{eq3.3} may be written as
\beq
\left<r^{2}\right>(r,\beta)\equiv r^{2}+\alpha(r,\beta)\frac{\hbar^{2}\beta}{m}
      \label{eq3.5}
\eeq
where the defined function $\alpha(r,\beta)$, multiplying the free particle
result, tends to 1 as $\beta\rightarrow 0$. $\alpha(0,\beta)$ is tabulated in 
Table III of the appendix.
Using Eq.~\ref{eq3.5} together with the
first order Wigner-Kirkwood quantum correction for the
free energy~\cite{landau2}, Eq.~\ref{eq3.2} gives the lowest order 
quantum correction to the enhancement factor
\beq
 h(0)[\Gamma,\eta]-h(0)[\Gamma,0]\approx
        -\frac{\eta}{4}[\alpha(0,\Gamma^{2}/\eta)- 1/2]\;.
                   \label{eq3.6}
\eeq
The result is shown for $\eta=0.1$ in Fig. 6 and qualitatively reproduces
the PIMC results showing significant reduction from the classical $h(0)$.
 
\section{Electron screening effects}

\begin{figure}[!]
 \includegraphics[angle=0,width=0.65\textwidth]{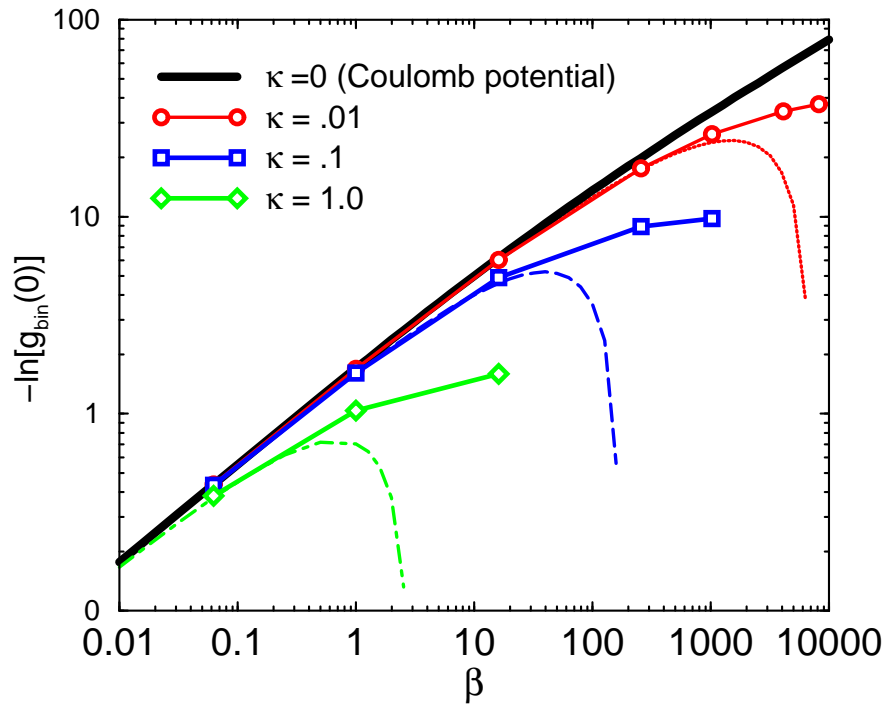}
\caption{ $g_{bin}(0)$ for the screened Coulomb potential at
          $\kappa =100$ (open circles), $10$ (open squares), and
          $1$ (open diamonds) compared to pure Coulomb result
         (thick solid line) and ``energy shift'' approximation
         $g^{\kappa}_{bin}(0)=e^{\beta\kappa} g^{\kappa=0}_{bin}(0)$.
            }
\label{FIG7}
\end{figure}

      The OCP model discussed above is justified when the electrons are sufficiently
degenerate to not respond to the ionic potential. Small deviations from this may
be treated by linear response leading to an effective ion-ion potential.
The most common such potential is the screened Coulomb potential
\beq
V(r)=\frac{e^{-\kappa r}}{r}
                               \label{eq4.1}
\eeq
with screening length $1/\kappa$, which we consider in this section.

   From the Feynman-Kac formula applied to the pair density matrix for $r=0$,
\beq
\begin{array}{c}\underline{\rho_{bin}(0,0;\beta)}\\\rho^{free}_{bin}(0,0;\beta)
\end{array}
       =\left<e^{-\int_{0}^{\beta}V[r(s)]ds}\right>_{BMP}
                                      \label{eq4.2}
\eeq
where the angular brackets denote an average over all Brownian motion paths
beginning and ending, after a time $\beta$, at the origin.
If the screened Coulomb potential is approximated as
\beq
\begin{array}{c}\underline{ e^{-\kappa r}}\\ r\end{array}
 \approx{ 1\over r}-\kappa
                   \label{eq4.3}
\eeq
 for $r\le1/\kappa$ then Eq.~\ref{eq4.2} becomes
\beq
\begin{array}{c}\underline{\rho_{bin}(0,0;\beta)}\\\rho^{free}_{bin}(0,0;\beta)
\end{array}
       =e^{\beta\kappa}\left<e^{-\int_{0}^{\beta}{1\over r(s)}ds}\right>_{BMP}
                                      \label{eq4.4}
\eeq
or $g_{bin}^{\kappa}(0)=e^{\beta\kappa}g_{bin}^{Coulomb}(0)$. The first factor may be
interpreted as an energy shift $\Delta E=-\kappa$. This
``constant energy shift'' approximation fails when the de Broglie thermal wavelength,
which gives a scale for the extent of the Brownian motion paths, is larger
than the screening length, $1/\kappa$, so the paths sample regions 
where Eq.~\ref{eq4.3} does not apply.

      Fig. 7 shows $g_{bin}(0)$ for several $\kappa$ values. These were computed here
by rewriting the Feynman-Kac formula as
\beq
\begin{array}{c}\underline{\rho_{bin}(0,0;\beta)}\\\rho^{\kappa =0}_{bin}(0,0;\beta)
\end{array}
       =\left<e^{-\int_{0}^{\beta}(V[r(s)]-1/r(s))ds}\right>_{CMP}
                                      \label{eq4.5}
\eeq
where now the difference between the screened and unscreened Coulomb
potential is integrated over paths distributed according to the 
unscreened Coulomb potential density matrix (CMP).

The dashed lines show the corresponding constant energy shift approximation.
This constant
energy shift approximation is valid in most
astrophysical applications~\cite{clayton} but, as seen in Fig. 7,
it can dramatically overestimate $g_{bin}(0)$ at low temperatures 
giving very misleading, excessive reaction rates~\cite{ichimaru}.

Turning now to many-body effects,
screening has been shown to reduce $h(0)$ in the classical OCP \cite{caillol}.
Its effect in the quantum OCP, shown in Fig. 8, is similar.
The reduced repulsion from surrounding
ions due to screening  again reduces the enhancement effect.

\begin{figure}[!]
\includegraphics[angle=0,width=0.55\textwidth]{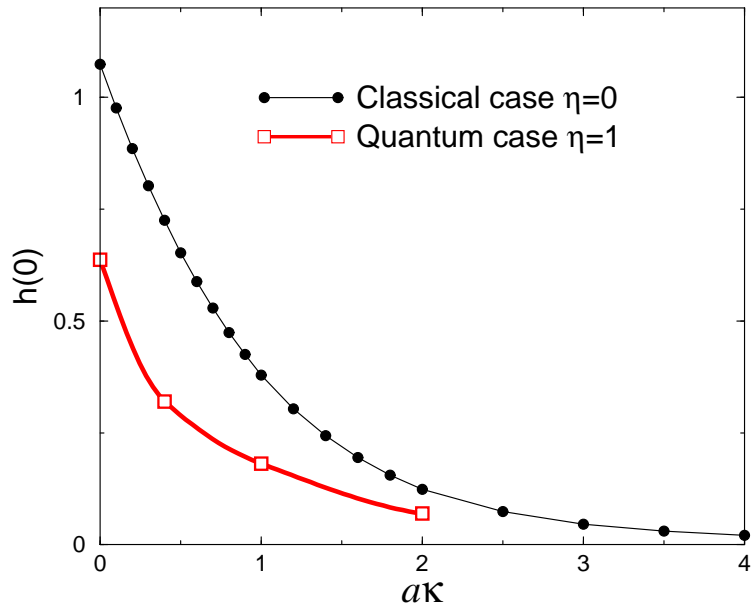}
\caption{ (color online) $h(0)$ for the classical~\cite{caillol}, $\eta=0$, and quantum,
$\eta=1$, Yukawa system at $\Gamma=40$ as a function of the inverse
screening length $\kappa$ times ion sphere radius $a$. $\kappa=0$
corresponds to the unscreened Coulomb potential results in Fig. 3.
            }
\label{FIG8}
\end{figure}


  

    

\section{ Quantum OCP thermodynamic properties}

      The PIMC calculations of the density matrix used to obtain $g(r)$ also yield
the kinetic and potential energies. Results for the computations used in this paper
are given in Table II.

       The PIMC and harmonic approximation values for the kinetic energy
\beq
      K={1\over 2}\sum_{k,j}\hbar \omega_{j}({\bf k})\left[
      \begin{array}{c}1\\ \overline{e^{\beta\hbar \omega_{j}({\bf k})}-1}\end{array}
      +{1\over 2}\right]
                                                    \label{eq5.1}
\eeq
where $\omega_{j}({\bf k})$ are the vibrational frequencies of the BCC Wigner lattice~\cite{kugler}
are compared in the table. Even though all simulations tabulated were in the fluid phase
the agreement at higher $\Gamma$ is quite good. This is not entirely surprising as similar
agreement with other properties has been often noted for the purely classical system
due to the long range nature of the interaction. The harmonic approximation also gives the
correct lowest order quantum correction,
\beq
    K={3\over 2}kT + {1\over 8} \eta^{2}/\Gamma
                                           \label{eq5.2}
\eeq
using the Kohn sum rule. As expected for fixed $\Gamma$ this agreement worsens as $\eta$
increases.

      The quantum corrections to the kinetic and potential energy as functions of 
$\Gamma$ and $\eta$ are plotted in figure 9. As expected the magnitude of the quantum corrections
for both quantities increases with $\eta$. At high $\Gamma$ the quantum kinetic and potential
energies are seen to converge. Again, if the harmonic approximation where theses quantum
corrections are equal, gives a good description of the thermodynamics even in the
liquid state at high $\Gamma$, then this convergence is understandable.
The slower convergence as $\eta$ increases for a fixed $\Gamma$ is also 
understandable since anharmonic corrections would be larger here.

\begin{figure}[!]
\includegraphics[angle=00,width=0.50\textwidth]{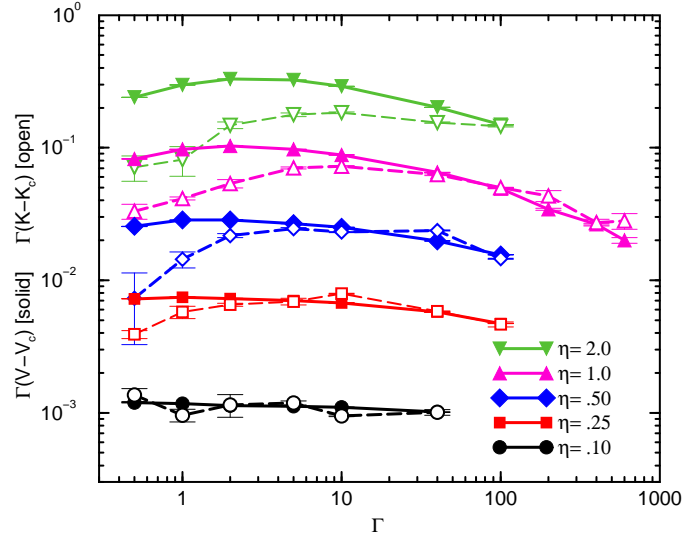}
\caption{ (color online) Comparison of the quantum kinetic, $K-K_{c}$ (open symbols 
          and dashed lines), and potential energy, $V-V_{c}$ (solid symbols and lines)
           per particle (data from Table II).
        }
\label{FIG9}
\end{figure}

      The quantum contributions to the total energy are compared to the expression for the second
order Wigner Kirkwood correction as given by Hansen and Viellefosse~\cite{hansen} in Table II
and in Fig. 10. Similar limits on the convergence of this second order approximation were
previously seen by Jones and Ceperley~\cite{jones}. (See their figure 3.)

\begin{figure}[!]
\includegraphics[angle=-90,width=0.45\textwidth]{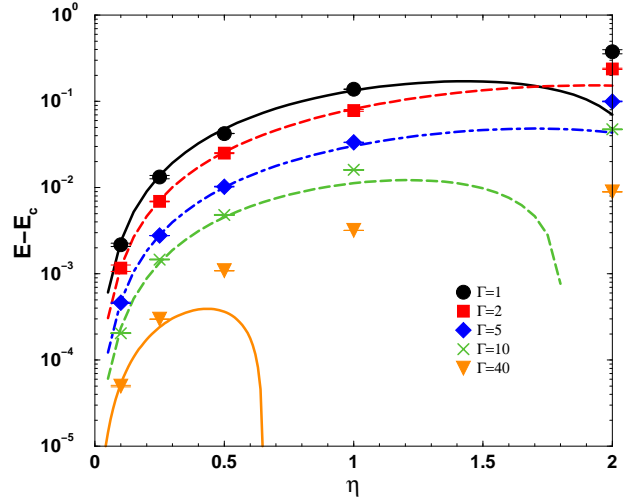}
\caption{ (color online) Comparison of the quantum energy per particle,
          $E-E_{c}$ for $\Gamma=1$ (solid circles), $\Gamma=2$ (solid squares)
          $\Gamma=5$, (solid diamonds), $\Gamma=10$ (crosses) and $\Gamma=40$ (solid triangles)
           with the second order Wigner-Kirkwood ($\hbar$ expansion) correction
          (corresponding solid or dashed lines).
        }
\label{FIG10}
\end{figure}

\begin{table}[!]
\caption[tab1]{Excess quantum kinetic ($K-K_{c}$), ``excess'' potential ($V-V_{c}$), 
       and internal ($E$) energy per particle from PIMC with 54 distinguishable 
	particles and $M$ time slices. $K_{c}=3/2\beta$ and $V_{c}$ is the potential 
       energy for the classical OCP with 54 particles in the periodic cell,
       taken as $\beta V_{c}=\{-0.24497, -0.57994,-1.32795,-3.7621,-7.9997,-34.2373\}$ for
       $\Gamma=\{0.5, 1, 2, 5, 10, 40\}$ 
       and as given by the fit$^{25}$ 
       $\beta V_{c}=-0.899375\Gamma+0.569333\Gamma^{1/3}-0.224470-0.017875/\Gamma^{1/3}$ for 
       larger $\Gamma$ values.
       The subscript $h$ denotes values from the harmonic approximation for the BCC Wigner lattice.
       The first and second order Wigner Kirkwood corrections to the energy from reference 25 are given in the last column
              }
\vspace*{.1in}
\begin{tabular}{|c|c|c|c|r|l|l|l|l|l|l|l|}\hline
$\eta$& $\Gamma$& $\beta$&$r_{s}$&  $M$ &   $~~K-K_{c}$    & $K_{h}-K_{c}$&     $~~~~~~~~V$        &  $~~V-V_{c}$  &    $~~~~~~~E$       &  $~~E-E_{c}$  &  ~~~~WK     \\\hline
0.1   &   0.5   & 2.5    & 5     &  25  &   0.0019(8)    & 0.0023        & -0.09557(3)    & 0.00239(3)   &  ~0.5063(8)  &  0.0043(8)   & ~0.00454  \\
0.1   &   1     & 10     & 10    &  10  &   0.0010(1)    & 0.0012        & -0.056824(5)   & 0.001170(5)  &  ~0.0941(1)  &  0.0022(1)   & ~0.00238 \\
0.1   &   2     & 40     & 20    &  40  &   0.0006(1)    & 0.00061       & -0.032631(4)   & 0.000567(4)  &  ~0.0054(1)  &  0.0012(1)   & ~0.00121  \\
0.1   &   5     & 250    & 50    &  25  &   0.000238(8)  & 0.000244      & -0.0148247(7)  & 0.0002235(7) &  -0.008587(8)&  0.000462(8) & ~0.000480\\
0.1   &   10    & 1000   & 100   &  10  &   0.000095(1)  & 0.00012       & -0.0078895(1)  & 0.0001102(1) &  -0.006295(1)&  0.000205(1) & ~0.000236\\
0.1   &   40    & 16000  & 400   &  80  &   0.000025(1)  & 0.000028      & -0.00211452(9  & 0.0000253(1) &  -0.001996(1)&  0.000050(1) & ~0.000053\\\hline
0.25  &   0.5   & 1.0    & 2     &  10  &   0.0078(5)    &  0.015        & -0.23046(3)    & 0.014493(3)  &  ~1.2773(5)  &  0.0223(5)   & ~0.02412  \\
0.25  &   1     & 4.0    & 4     &  40  &   0.0058(6)    & 0.0077        & -0.13753(3)    & 0.00745(3)   &  ~0.2432(6)  &  0.0133(6)   & ~0.01381  \\
0.25  &   2     & 16     & 8     &  16  &   0.00327(9)   & 0.00382       & -0.079355(4)   & 0.003642(4)  &  ~0.01766(8) &  0.00691(8)  & ~0.00713  \\
0.25  &   5     & 100    & 10    &  80  &   0.00138(8)   & 0.0015        & -0.036220(4)   & 0.001401(4)  &  -0.01984(8) &  0.00278(8)  & ~0.002819\\
0.25  &   10    & 400    & 40    &  40  &   0.000791(7)  & 0.000727      & -0.0193250(6)  & 0.0006743(6) &  -0.014784(7)&  0.001465(7) & ~0.001348\\
0.25  &   40 	& 6400   & 160   &  80  &   0.000146(2)  & 0.000156      & -0.0052058(1)  & 0.0001438(1) &  -0.004826(2)&  0.000290(2) & ~0.000241\\
0.25  &   100   & 40000  & 400   &  200 &   0.000047(2)  & 0.000051      & -0.00213982(9) & 0.0000472(1) &  -0.002056(2)&  0.000094(2) & ~0.000022\\\hline
0.50  &   0.5   & 0.5    & 1     &  50  &   0.015(8)     & 0.016         & -0.4389(1)     & 0.05099(1)   &  ~2.576(8)   &  0.066(8)    & ~0.06798 \\
0.50  &   1     & 2.0    & 2     &  50  &   0.014(2)     & 0.0306        & -0.26155(4)    & 0.02842(4)   &  ~0.503(2)   &  0.0424(2)   & ~0.04797  \\
0.50  &   2     & 8.0    & 4     &  40  &   0.0109(4)    & 0.0151        & -0.15176(1)    & 0.01423(1)   &  ~0.0466(4)  &  0.0251(4)   & ~0.02582  \\
0.50  &   5     & 50     & 10    &  50  &   0.00491(9)   & 0.00582       & -0.069885(4)   & 0.005356(4)  &  -0.03497(9) &  0.01027(9)  & ~0.01005 \\
0.50  &   10    & 200    & 20    &  50  &   0.00231(2)   & 0.00274       & -0.037492(2)   & 0.002506(2)  &  -0.02768(2) &  0.00482(2)  & ~0.00453   \\
0.50  &   40    & 3200   & 80    &  80  &   0.000593(5)  & 0.000541      & -0.0102038(6)  & 0.0004954(6) &  -0.009142(5)&  0.001088(5) & ~0.000364\\
0.50  &   100   & 20000  & 200   &  100 &   0.000145(2)  & 0.000166      & -0.0042185(2)  & 0.0001555(2) &  -0.003998(2)&  0.000301(2) & -0.000452\\\hline
1.0   &   0.5   & 0.25   & 0.5   &  25  &   0.066(9)     & 0.245         & -0.8154(2)     & 0.1644(2)    &  ~5.251(9)   &  0.230(9)    & ~0.04381  \\
1.0   &   1     & 1.0    & 1     &  10  &   0.041(1)     & 0.121         & -0.48277(6)    & 0.09722(6)   &  ~1.059(1)   &  0.138(1)    & ~0.1338   \\
1.0   &   2     & 4.0    & 2     &  40  &   0.027(2)     & 0.059         & -0.28043(5)    & 0.05156(5)   &  ~0.121(2)   &  0.079(2)    & ~0.0816   \\
1.0   &   5     & 25     & 5     &  25  &   0.0141(2)    & 0.0219        & -0.13102(2)    & 0.01947(2)   &  -0.0569(2)  &  0.0336(2)   & ~0.03044  \\
1.0   &   10    & 100    & 10    &  80  &   0.00723(9)   & 0.0100        & -0.071170(6)   & 0.008827(4)  &  -0.04894(9) &  0.01606(9)  & ~0.0112   \\
1.0   &   40    & 1600   & 40    &  128 &   0.00156(2)   & 0.00178       & -0.019772(2)   & 0.0016270(7) &  -0.01727(2) &  0.00319(3)  & -0.00334\\
1.0   &   100   & 10000  & 100   &  100 &   0.000495(2)  & 0.00052       & -0.008258(2)   & 0.0004905(7) &  -0.007606(2)&  0.000986(3) & -0.00611\\
1.0   &   200   & 40000  & 200   &  400 &   0.00021(2)   & 0.000198      & -0.004248(3)   & 0.000172(3)  &  -0.00399(2) &  0.00038(2)  & -0.00697 \\
1.0   &   400   & 160000 & 400   &  400 &   0.000068(4)  & 0.000074      & -0.0021575(9)  & 0.0000662(9) &  -0.002080(4)&  0.000134(4) & -0.00735\\
1.0   &   600   & 360000 & 600   &  450 &   0.000047(6)  & 0.000041      & -0.001453(2)   & 0.0000334(2) &  -0.001402(7)&  0.000080(7) & -0.00746 \\\hline
2.0   &   0.5   & 0.125  & 0.25  &  50  &   0.14(3)      & 0.97          & -1.4797(4)     & 0.4799(4)    &  10.66(3)    &  0.62(3)     & -1.6495 \\
2.0   &   1     & 0.5    & 0.5   &  50  &   0.08(2)      & 0.471         & -0.8625(3)     & 0.2974(3)    &  ~2.22(2)    &  0.38(2)     & ~0.07017  \\
2.0   &   2     & 2.0    & 1     &  50  &   0.074(4)     & 0.224         & -0.4993(1)     & 0.1647(1)    &  ~0.325(4)   &  0.239(4)    & ~0.1526   \\
2.0   &   5     & 12.5   & 2.5   &  50  &   0.035(1)     & 0.080         & -0.23622(5)    & 0.06475(5)   &  -0.081(1)   &  0.010(1)    & ~0.04349  \\
2.0   &   10    & 50     & 5     &  50  &   0.0184(3)    & 0.0345        & -0.13097(2)    & 0.02902(2)   &  -0.0826(3)  &  0.0474(3)   & -0.01006\\
2.0   &   40    & 800    & 20    &  80  &   0.00387(5)   & 0.00565       & -0.037748(4)   & 0.005049(4)  &  -0.03201(5) &  0.00892(5)  & -0.0517\\
2.0   &   100   & 5000   & 50    &  100 &   0.00145(2)   & 0.00158       & -0.016003(2)   & 0.001493(2)  &  -0.01425(1) &  0.00294(1)  & -0.0589\\\hline
\end{tabular}
\end{table}

\newpage

\section{Conclusions}

In conclusion, quantum effects have been shown to significantly reduce
the many-body enhancement factor which influences nuclear reaction rates
in dense plasmas. Electron screening effects produce a further reduction.
The contact probability and reaction rates based on it increase
monotonically with temperature and density.
The relation between the contact probability and free energy
differences is derived and intuitively illustrated. 

\appendix
\section{Small $r$ expansion of $h(r)$ in the semiclassical limit}

      In a classical plasma  $g(0)$ can be related to a
free energy difference \cite{graboske} and a simple, explicit value 
given for the coefficient of the  $r^{2}$ expansion term \cite{jancovici}. For
the quantum plasma no similar, exact relation has been found however an 
expansion in the ratio of the de Broglie thermal wavelength to the ion 
sphere radius for $g(r)$ at small $r$ can be made \cite{jancovici}.
This expansion, which reduces to the classical result, is reviewed here.
The principal results are given below in Eqs.~\ref{eqa12},~\ref{eqa17},~\ref{eqa18} 
and in Table III for $\alpha(0;\beta)$.

     Starting from the definition of the radial distribution function
for a one component system
\beq
g({\bf r})= \Omega\left<\delta({\bf r}-{\bf r_{12}})\right> 
          \equiv \Omega\begin{array}{c}\underline{ Tr\left[e^{-\beta H}
   \delta({\bf r}-{\bf r_{12}})\right]}\\ 
   Tr\left[e^{-\beta H}\right] \end{array}=
\Omega\begin{array}{c}\underline{ Tr\left[e^{-\beta H}
   \delta({\bf r}-{\bf r_{12}})\right]}\\ {\cal Z}\end{array}
                                                         \label{eqa1}
\eeq
where particles one and two have been singled out, the Hamiltonian
\beq
H=\sum_{j=1}^{N}K_{j} +\sum_{k}\sum_{j<k}v_{jk}
                                                   \label{eqa2}
\eeq
with $K$ the kinetic energy operator is rewritten in terms of 
center of mass and relative coordinates for the "reacting" particles,
one and two, ${\bf R}=({\bf r_{1}}+{\bf r_{2}})/2\;$, 
${\bf r_{12}}={\bf r_{1}}-{\bf r_{2}}$. Doing this, separating out
the terms involving only particles one and two, and adding and 
subtracting a term representing the interaction of both particles one
and two at the center of mass position with the "spectator"
particles $3\ldots N$ the Hamiltonian can be rewritten as,
\beq
H = H_{rel}({\bf r}) +H_{mix}({\bf R},{\bf r}_{3}\ldots {\bf r}_{N}) +
    \Delta V({\bf R},{\bf r},{\bf r}_{3}\ldots {\bf r}_{N})\;.
                                                   \label{eqa3}
\eeq
The "relative" Hamiltonian
\beq
 H_{rel}({\bf r}) = K_{r} + v({\bf r})
                                                   \label{eqa4}
\eeq
commutes with the "mixture" Hamiltonian
\beq
H_{mix}({\bf R},{\bf r}_{3}\ldots {\bf r}_{N})= K_{R}+\sum_{j=3}^{N}K_{j} 
   +\sum_{j=3}^{N}\sum_{k>j} v_{jk} +2\sum_{j=3}^{N}v({\bf R}-{\bf r_{j}})
                                                   \label{eqa5}
\eeq
which corresponds to a particle of double the mass and charge at
the center of mass position, ${\bf R}$,  and the $N-2$ spectator particles.
The coupling term
\beq
\begin{array}{ll}
\Delta V({\bf R},{\bf r},{\bf r}_{3}\ldots {\bf r}_{N})&= \sum_{j=3}^{N} \left[
v({\bf r_{1}}-{\bf r_{j}}) + v({\bf r_{2}}-{\bf r_{j}})-2v({\bf R}-{\bf r_{j}})
   \right]\\
   &=  \sum_{j=3}^{N}\left[ v( {\bf R+\frac{r}{2}} - {\bf r_{j}} ) + 
     v( {\bf R-\frac{r}{2}} - {\bf r_{j}} ) -2 v( {\bf R}-{\bf r_{j}} )
     \right]\\
 &= \sum_{j=3}^{N}{\bf \nabla}{\bf \nabla}v({\bf R}-{\bf r_{j}}):
     {\bf r} {\bf r}/4 + O({\bf r}^{4})
\end{array}
                                                   \label{eqa6}
\eeq
is the difference between the interactions of all other particles in the system
with  particles 1 and 2 at their actual positions minus these interactions 
when particles 1 and 2 are fused at their center of mass position.

 Using Eq.~\ref{eqa3} in the expression for $g(r)$ and 
taking the trace in real space
\beq
g(r) = \begin{array}{c}\underline{\Omega }\\{\cal{Z}}\end{array}
\int \left<{\bf R},{\bf r},{\bf r_{3}},\ldots {\bf r_{N}}|
e^{-\beta(H_{rel}+H_{mix}+\Delta V)}|
      {\bf R},{\bf r},{\bf r_{3}},\ldots {\bf r_{N}}\right>
      d{\bf R}d{\bf r_{3}}\ldots d{\bf r_{N}}  \;.
                                                   \label{eqa7}
\eeq
When averaged over an isotropic system
\beq
{\bf \nabla}{\bf \nabla} v({\bf R}-{\bf r_{j}})=
\nabla^{2} v({\bf R}-{\bf r_{j}})\stackrel{\leftrightarrow}{\bf I}\;.
                                                   \label{eqa8}
\eeq
The Laplacian is easily evaluated for the Coulomb system considered here
where the interaction, accounting for periodic boundary conditions and
charge neutrality, is the Ewald potential, $\Psi_{Ewald}(r)$.
Using $\nabla^{2}\Psi_{Ewald}(r) = -4\pi\delta(r)+4\pi/\Omega$,
where, physically, the constant comes from the neutralizing background,
\beq
\Delta V({\bf R},{\bf r},{\bf r}_{3}\ldots {\bf r}_{N})= 
     \frac{4\pi Z^{2}(N-2)}{\Omega}\frac{r^{2}}{12}+O(r^{4})
  = \frac{Z^{2}r^{2}}{4 a^{3}}+O(r^{4})\;.
                                           \label{eqa9}
\eeq
The simplification that the coefficient of the lowest order term in $r$
is constant is unique to Coulomb systems.

   With this expansion and using the fact that $H_{mix}$ commutes with
$H_{rel}$ and the above lowest order term for $\Delta V$ 
\beq
\begin{array}{ll}
 g(r) &= \begin{array}{c}\underline{\Omega }\\{\cal{Z}}\end{array}
  \left< {\bf r}|e^{-\beta(H_{rel}+Cr^{2})}|{\bf r} \right>
 \int \left< {\bf R},{\bf r_{3}},\ldots {\bf r_{N}}|
 e^{-\beta H_{mix}}|
      {\bf R},{\bf r_{3}},\ldots {\bf r_{N}} \right>
  d{\bf R}d{\bf r_{3}}\ldots d{\bf r_{N}} +O(<r^{4}>)\\
 \nonumber
  &= \Omega e^{-\beta(F_{mix}-F_{pure})}
  \left< {\bf r}|e^{-\beta(H_{rel}+Cr^{2})}|{\bf r} \right>+O(<r^{4}>)
        \label{eqa10}
\end{array}
\eeq
where $C=Z^{2}/4 a^{3}$. The free energies  $F_{pure}$, ${\cal Z}=\exp(-\beta F_{pure})$,
 and 
$F_{mix}$,
\beq
  e^{-\beta F_{mix}}=\int \left< {\bf R},{\bf r_{3}},\ldots {\bf r_{N}}|
                 e^{-\beta H_{mix}}| {\bf R},{\bf r_{3}},\ldots {\bf r_{N}} \right>
  d{\bf R}d{\bf r_{3}}\ldots d{\bf r_{N}}
\eeq
correspond to the original, total Hamiltonian and to $H_{mix}$ 
respectively. They are the fully quantum mechanical free energies.

     Introducing the binary radial distribution function, from the definition
Eq.~\ref{eqa1} applied to a two particle system,
\beq
g_{bin}(r)=\left(4\pi\hbar^{2}\beta/m\right)^{2}
       \left< {\bf r}|e^{-\beta H_{rel}}|{\bf r} \right>
            \label{eqa11}
\eeq
the expansion Eq.~\ref{eqa10} for $g(r)$ becomes
\beq
g(r) =g_{bin}(r) \Omega \left(\frac{m}{4\pi\hbar^{2}\beta}\right)^{3/2}
                 e^{-\beta(F_{mix}-F_{pure})}\begin{array}{c}
\underline{ \left< {\bf r}|e^{-\beta(H_{rel}+C r_{12}^{2})}|{\bf r} \right>}\\
\left< {\bf r}|e^{-\beta H_{rel}}|{\bf r} \right> \end{array}+O(\left<r^{4}\right>)\;.
     \label{eqa12}
\eeq
The term following $g_{bin}$ cancels the remaining ``ideal gas term''
in the difference between the $N$ particle $F_{pure}$ and the $N-1$ particle
$F_{mix}$. In the following this term is omitted and the free energies
refer only to the nonideal terms.

     For the classical case $H_{rel}$ and $Cr^{2}$ commute and the
last ratio in equation above is just $e^{-Cr^{2}}$ so
\beq
g(r) =g_{bin}(r)e^{-\beta(F_{mix}-F_{pure} - Cr^{2})}
     =g_{bin}(r)e^{-\beta(F_{mix}-F_{pure})-\frac{\Gamma}{4} (r/a)^{2}} \;.
\eeq
This relation, with the correct $r^{2}$ term, was first derived in reference 16
correcting an earlier factor of two error in reference 2.

   As a simple application consider the weak coupling, Debye-H{\"u}ckel
limit. Using the well known interaction free energy for this model \cite{landau}
\beq
    F= -\frac{2}{3}\sqrt{\frac{\pi}{TV}}\left(\sum_{s}N_{s}Z_{s}^{2}\right)^{3/2}
\eeq 
applied to the "pure" ($N_{1}=N\;$, $Z_{1}=Z$) and the "mixture" case
($N_{1}=N-2,\;$ $Z_{1}=Z,\;$ $N_{2}=1,\;$ $Z_{2}=2Z$) gives
$h(0)=\ln[g(0)/g_{bin}(0)]=\sqrt{3\Gamma}$. This result also comes from expanding
the screened potential form, $g(r)=e^{-\kappa r}/r$ with $\kappa$ the inverse
Debye-H{\"u}ckel screening length. The next term in the expansion, linear in $r$, is seen 
to be incorrect however. 

     In the quantum mechanical case the terms $H_{rel}$ and $Cr^{2}$
 do not commute. The identity
\begin{eqnarray}
e^{-\beta(H_{rel}+Cr^{2})}=e^{-\beta H_{rel}}-\int_{0}^{\beta}
 e^{-(\beta-s) H_{rel}} Cr^{2} e^{-s (H_{rel}+Cr^{2})}ds\\
=e^{-\beta H_{rel}}-\int_{0}^{\beta}e^{-(\beta-s) H_{rel}} Cr^{2} 
                                    e^{-s H_{rel}}ds +O(r^{4})
\end{eqnarray}
leads to
\beq
 \begin{array}{c} 
   \underline{ \left< {\bf r}|e^{-\beta(H_{rel}+C r^{2})}|{\bf r} \right>}\\
    \left< {\bf r}|e^{-\beta H_{rel}}|{\bf r} \right> \end{array}=
1-\beta C<r^{2}>+O(<r^{4}>)
             \label{eqa17}
\eeq
where
\beq
\left<r^{2}\right>\equiv \frac{1}{\beta}\int_{0}^{\beta}ds\int d{\bf r}'
   \rho_{rel}({\bf r},{\bf r}';\beta-s) r'^{2} \rho_{rel}({\bf r}',{\bf r};s)/
\rho_{rel}({\bf r},{\bf r};\beta)\;.
                                                 \label{eqa18}
\eeq
Unlike the classical case $<r^{2}>\neq r^{2}$ but, because of the range of
$\rho_{rel}$, will differ from it by an amount proportional to the
squared de Broglie thermal wavelength. For example, if $\rho_{rel}$ is
approximated by only the free particle term then
\beq
<r^{2}>=r^{2}+\hbar^{2}\beta/M\;.
\eeq
When $r\rightarrow 0$, $<r^{2}>$
and higher order terms are 
thus nonzero and the simple, classical  relation between the screening function 
at $r=0$ and the free energy difference no longer applies. What has been
generated is a double expansion in $r^{2}$ and the squared ratio of the
de Broglie thermal wavelength to the ion sphere radius.

      The lowest order quantum correction to the free energy \cite{landau2}
\beq
\beta F=\beta F_{classical}+\frac{\hbar^{2}\beta^{2}}{24}\left<
  \sum_{j}\begin{array}{c}\underline{{\bf \nabla}^{2}_{j}U}\\M_{j}\end{array}
\right>
\eeq
where $U$ is the total potential energy function, applied to the uniform
background model and following the algebra leading to equation \ref{eqa9},
gives
\beq
\beta\left(F_{mix}-F_{pure}\right)= 
    \beta\left(F_{mix}^{classical}-F_{pure}^{classical}\right)-\frac{1}{8}\eta
     \Gamma   \;.
\eeq
The quantum correction in this term would further increase the enhancement
factor contrary to what is found. 

     The $\beta C<r^{2}>$ term in Eq.~\ref{eqa17}  corrects this. 
We have not found a simple expression for $<r^{2}>$ in terms of
continuum Coulomb wave functions however
it is not difficult to evaluate numerically using
the axial symmetry of the ${\bf r}'$ integral (or radial symmetry when
${\bf r}=0$). The result can be expressed
as
\beq
\left<r^{2}\right>(r,\beta)\equiv r^{2}+\alpha(r,\beta)\frac{\hbar^{2}\beta}{M}
                                                  \label{eqa19}
\eeq
where the function multiplying the free particle result,
$\alpha(r,\beta)\rightarrow 1$ as $\beta\rightarrow 0$. As $\beta$ increases
$\alpha$ slowly increases, reflecting the tendency of the repulsive Coulomb
potential to emphasize  larger radius ''paths'' compared to the 
free particle limit.

Adding this term to the free energy change, the lowest order
quantum correction in the enhancement factor is
\beq
\Delta h_{QM}(0)=-\frac{\eta}{4}[\alpha(0,\Gamma^{2}/\eta)-1/2]\;.
\eeq
This now correctly predicts the decrease in $h(0)$ due to quantum
effects. 
 $\alpha(0,\beta)$ is tabulated in table III.
This semiclassical expansion was compared with the PIMC results for the
case $\eta=0.1$ in Fig.~\ref{FIG6}.
For values of $\eta\geq 0.25$ the lowest order expansion overestimates the quantum effects
by almost a factor of 2.

\vspace*{.1in}
\begin{table}[!]
\caption[tab3]{
                $\alpha(0,\beta)$ as defined in Eq.~\ref{eqa19} and \ref{eqa18}.

               }
\begin{tabular}{|c|c|}\hline
 $\beta$  &  $\alpha(0,\beta)$ \\\hline
0.0  & 1.0   \\
0.5  & 1.106   \\
1.0  & 1.148   \\
2.0  & 1.206   \\
4.0  & 1.286  \\
6.0  & 1.345   \\
8.0  & 1.394   \\
10.0  & 1.435   \\ 
20.0  & 1.591   \\
30.0  & 1.703   \\\hline
\end{tabular}
\end{table}

\vspace*{3mm}
We thank H. DeWitt for numerous conversations
and for several of the classical $h(0)$ values displayed in figure 3.
This work was performed under the auspices
of the U.S. Department of Energy by University of California LLNL
contract \mbox{No. W-7405-Eng-48} in a CREM free environment.

\end{document}